\documentclass[5p,sort&compress]{elsarticle}

\usepackage{hyperref}
\usepackage{amsmath}
\usepackage{braket}
\usepackage{SIunits}
\usepackage{miller}
\usepackage{textcomp} 
\usepackage{todonotes}

\renewcommand{\vec}[1]{\ensuremath{\boldsymbol{#1}}}
\newcommand{\op}[1]{\ensuremath{\hat{#1}}}
\newcommand{\vecop}[1]{\ensuremath{\hat{\vec{#1}}}}
\newcommand{\matop}[1]{\ensuremath{\hat{\mat{#1}}}}
\newcommand{\mat}[1]{\ensuremath{\boldsymbol{#1}}}
\newcommand{\ketbra}[2]{\ensuremath{\ket{#1}\bra{#1}}}

\newcommand{\ee}{\ensuremath{\mathrm{e}}}
\newcommand{\dd}{\ensuremath{\mathrm{d}}}
\newcommand{\ii}{\ensuremath{\mathrm{i}}}

\newcommand{\wigner}[6]{\ensuremath{\begin{pmatrix} #1 & #3 & #5 \\ #2 & #4 & #6 \end{pmatrix}}}

\begin{document}

\title{A pure state decomposition approach of the mixed dynamic form factor for mapping atomic orbitals}
\author[ifp]{S. L\"offler}
\ead{stefan.loeffler@tuwien.ac.at}
\author[ifp]{V. Motsch}
\author[ifp,ustem]{P. Schattschneider}

\address[ifp]{Institute of Solid State Physics, Vienna University of Technology, Vienna, Austria}
\address[ustem]{University Service Centre for Electron Microscopy, Vienna University of Technology, Vienna, Austria}

\begin{abstract}
We demonstrate how the mixed dynamic form factor (MDFF) can be interpreted as a quadratic form. This makes it possible to use matrix diagonalization methods to reduce the number of terms that need to be taken into account when calculating the inelastic scattering of electrons in a crystal. It also leads in a natural way to a new basis that helps elucidate the underlying physics. The new method is applied to several cases to show its versatility. In particular, predictions are made for directly imaging atomic orbitals in crystals.
\end{abstract}

\begin{keyword}
inelastic scattering \sep mixed dynamic form factor \sep orbital mapping \sep EFTEM
\end{keyword}

\maketitle

\section{Introduction}

Nowadays, simulations are indispensable both for planning and for interpreting experiments in the transmission electron microscope (TEM), in particular when working with electron energy loss spectrometry (EELS). The key quantity for simulating inelastic electron scattering is the mixed dynamic form factor (MDFF) \cite{U_v15_i3_p173,Kohl1985,PRB_v59_i16_p10959,M_v31_i4_p333}. In many cases, this complex quantity is simplified by several approximations, like, for instance, the dipole approximation. Recently, it has been shown, however, that this can lead to quite severe errors \cite{U_v111_i_p1163}. Furthermore, with recent advances of aberration corrected microscopes, more accurate calculations of the MDFF will become essential for future experiments.

In this work, we will give a brief repetition of the mixed dynamic form factor. It has been well known for a long time that in dipole approximation, the MDFF can be written in the form $a \vec{q} \cdot \vec{q}'$ ($+ \vec{q} \times \vec{q}'$ in the case of magnetism; see, e.g., \cite{PhysRevB.54.3861,U_v108_i5_p433}). Our work goes beyond this approximation by showing that all multipole orders can be written as a quadratic form. This is followed by an analysis of how a basis transformation can bring it into a simpler, diagonal form that is much easier to handle numerically. Furthermore, the physical significance of this procedure will be outlined. The general concept of factorizing and diagonalizing density matrices (i.e., writing the corresponding density operator as an incoherent sum of pure states) is well known \cite{Blum1996,TJoCP_v37_i3_p577} and is also applied in other fields (e.g., \cite{JoMO_v57_i13_p1109}). However, to the best of our knowledge, it was not yet applied in the way presented here to simplify the MDFF.

In the last part, the new formalism will be applied to both existing and new measurement setups to study its applicability and versatility.

\section{The mixed dynamic form factor and its pure state decomposition}

In the most general approach, the quantum mechanical system consisting of both the probe electron and the sample can best be described by a density operator $\matop{\rho}$ or its matrix elements, the so-called density matrix $\mat{\rho}$ \cite{Blum1996}. Adopting the density matrix formalism instead of the simpler wave function approach is greatly beneficial as one cannot observe the target's final state directly. This ignorance of a part of the system after an inelastic interaction gives rise to a mixed state which can be described very effectively using the density matrix \cite{M_v31_i4_p333,Blum1996,PRB_v59_i16_p10959}.

Before the interaction, the probe and the target systems can be considered independent. For the sake of simplicity, we will furthermore assume that both systems are initially in a pure state, i.e., each can be described by a single wave function. Then, the density operator of the whole system before the interaction is given by
\begin{equation}
	\matop{\rho}_{\text{tot},0} = \ketbra{i}{i} \otimes \ketbra{I}{I} = \ket{I} \ket{i}\bra{i}\bra{I},
\end{equation}
where $\otimes$ denotes the direct product. Throughout this article, we use small letters when referring to the probe beam and capital letters when referring to the target.

In first order Born approximation, the density operator after the inelastic interaction mediated by an interaction potential $\op{V}$ is given by
\begin{equation}
	\matop{\rho}_\text{tot}= \op{V}\ket{I} \ket{i}\bra{i}\bra{I} \op{V}^\dag \delta(E + E_I - E_F),
\end{equation}
where $E$ is the ``energy loss'' of the probe beam (i.e., the energy transfered from the probe beam to the target electron), and $E_I, E_F$ are the initial and final state energies of the target. Since the target system is not observed directly, one has to construct the reduced density operator for the probe beam by summing incoherently over all possible final states of the target. This reduced density operator is given by
\begin{equation}
	\matop{\rho}= \sum_{F} \braket{F | \op{V} |I} \ket{i}\bra{i}\braket{I | \op{V}^\dag |F} \delta(E + E_I - E_F),
\end{equation}
which can then be propagated elastically through the crystal and be used to predict the outcome of measurements in different geometries. It must be emphasized that the ordering of the terms is vital here, since $\op{V}$ in general acts on both the probe and the target states, which results in an entanglement of the two.

In EELS experiments, the interaction operator $\op{V}$ is the Coulomb interaction operator. Its two most common basis representations are in configuration space,
\begin{equation}
\begin{aligned}
	\op{V}(\vec{r}, \vec{r}') = \braket{\vec{r} | \op{V} | \vec{r}'} &= \frac{e^2}{4\pi\epsilon_0} \frac{\delta(\vecop{R} - \vecop{R}') }{|\vec{r} - \vecop{R}|} \delta(\vec{r} - \vec{r}') \\
	&=: \op{V}(\vec{r}) \delta(\vec{r} - \vec{r}'),
\end{aligned}
\end{equation}
and in reciprocal space,
\begin{equation}
\begin{aligned}
	\op{V}(\vec{k}, \vec{k}') = \braket{\vec{k} | \op{V} | \vec{k}'} &= 
	\frac{e^2}{4\pi\epsilon_0} \frac{\ee^{\ii (\vec{k}'-\vec{k}) \cdot \vecop{R}}}{|\vec{k}'-\vec{k}|^2} \delta(\vecop{R} - \vecop{R}') \\
	&=: \frac{e^2}{4\pi\epsilon_0} \frac{\ee^{\ii \vec{q} \cdot \vecop{R}}}{|\vec{q}|^2} \delta(\vecop{R} - \vecop{R}') \\
	&=: \op{V}(\vec{q}).
\end{aligned}
\end{equation}
Here, $e$ is the elementary charge and $\epsilon_0$ is the permittivity.

In these two representations, the reduced density matrix reads
\begin{equation}
\begin{aligned}
	\mat{\rho}(\vec{r}, \vec{r'})
	=& -4\pi^2  \sum_{F} \int \dd\vec{\tilde{r}} \bra{F} \braket{\vec{r} | \op{V} | \vec{\tilde{r}}} \ket{I} \braket{\vec{\tilde{r}} | i} \\
	& \int \dd\vec{\tilde{r}}' \braket{i | \vec{\tilde{r}}'} \bra{I} \braket{\vec{\tilde{r}}' | \op{V}^\dag | \vec{r}'} \ket{F} \delta(E+E_I-E_F) \\
	=& -4\pi^2 \sum_{F} \braket{F | \op{V}(\vec{r}) | I}\braket{I | \op{V}^\dag(\vec{r}') | F} \\
	& \braket{\vec{r} | i} \braket{i | \vec{r}'} \delta(E+E_I-E_F) \\
	=& S(\vec{r}, \vec{r}') \braket{\vec{r} | i} \braket{i | \vec{r}'} \\
	\mat{\rho}(\vec{k}, \vec{k'})
	=& -4\pi^2\sum_{F} \int \dd\vec{\tilde{k}} \bra{F} \braket{\vec{k} | \op{V} | \vec{\tilde{k}}} \ket{I} \braket{\vec{\tilde{k}} | i} \\
	& \int \dd\vec{\tilde{k}}' \braket{i | \vec{\tilde{k}}'} \bra{I} \braket{\vec{\tilde{k}}' | \op{V}^\dag | \vec{k}'} \ket{F} \delta(E+E_I-E_F) \\
	=& -4\pi^2 \sum_{F} \iint \dd\vec{q}\dd\vec{q}' \braket{F | \op{V}(\vec{q}) | I} 
	\braket{I | \op{V}^\dag(\vec{q}') | F} \\
	& \braket{\vec{k} + \vec{q} | i} \braket{i | \vec{k}' + \vec{q}'} \delta(E+E_I-E_F) \\
	=& \iint \dd\vec{q}\dd\vec{q}' S(\vec{q}, \vec{q}') \braket{\vec{k} + \vec{q} | i} \braket{i | \vec{k}' + \vec{q}'}.
\end{aligned}
\end{equation}
Here, the MDFF $S(\vec{q}, \vec{q}')$ as well as the real-space MDFF (rMDFF) $S(\vec{r}, \vec{r}')$ were introduced which are related by a Fourier transformation.\footnote{Contrary to the convention adopted in previous works, we include the $1/{q}^2{q'}^2$ term in the definition of the MDFF as it makes the definition more concise and easy to use.} It is noteworthy that --- due to the particular properties of the Coulomb operator --- the rMDFF can be multiplied on the initial probe wave functions, whereas the MDFF has to be convolved with them.

In order to perform calculations, one not only has to specify a basis for the probe states, but also for the target states. Usually, one chooses a spherical harmonics basis which is particularly useful for describing the tightly bound initial states that give rise to EELS core losses. Hence, the initial state is written as $\ket{l\frac{1}{2}jj_z}$\footnote{This takes into account the spin-orbit coupling of the tightly bound core states \cite{N_v441_i_p486}.}, while the final states are expanded in terms of $\ket{LM\frac{1}{2}S}$. In the following, we will also sum incoherently over $j_z$ since that quantum number of the initial state is typically unknown. In the Kohn-Sham approximation, the MDFF is then given by~\cite{N_v441_i_p486,PRB_v75_i21_p214425,Nelhiebel1999}
\begin{equation}
\begin{aligned}
	S(\vec{q}, \vec{q}') 
	=& -4\pi^2 \sum_{Fj_z} \sum_{LMSL'M'S'} \delta(E+E_I-E_F) \\
	& \braket{F | LM\frac{1}{2}S} \braket{LM\frac{1}{2}S | \op{V}(\vec{q}) | l\frac{1}{2}jj_z } \\
	& \braket{l\frac{1}{2}jj_z | \op{V}^\dag(\vec{q}') | L'M'\frac{1}{2}S'} \braket{L'M'\frac{1}{2}S' | F}  \\
	=& -\frac{\pi e^4}{\epsilon_0^2 {q}^2{q'}^2} (2l+1) (2j+1) \sum_{mm'}\sum_{LMS}\sum_{L'M'S'} \sum_{\lambda\mu\lambda'\mu'} \\
	& \ii^{\lambda-\lambda'} \sqrt{(2\lambda+1)(2\lambda'+1)(2L+1)(2L'+1)} \\
	& Y_{\lambda}^{\mu}(\vec{q}/q)^* \langle j_{\lambda}(q) \rangle_{ELSj} Y_{\lambda'}^{\mu'}(\vec{q}'/q') \langle j_{\lambda'}(q') \rangle_{EL'S'j} \\
	& \wigner{l}{0}{\lambda}{0}{L}{0} \wigner{l}{0}{\lambda'}{0}{L'}{0} \\
	& \wigner{l}{-m}{\lambda}{\mu}{L}{M} \wigner{l}{-m'}{\lambda'}{\mu'}{L'}{M'} \\
	& \sum_{j_z} (-1)^{m+m'} \wigner{l}{m}{\frac{1}{2}}{S}{j}{-j_z} \wigner{l}{m'}{\frac{1}{2}}{S'}{j}{-j_z} \\
	& \sum_{\vec{k}n} D_{LMS}^{\vec{k}n} \left( D_{L'M'S'}^{\vec{k}n} \right)^* \delta(E + E_{nlj} - E_{\vec{k}n}).
\end{aligned}
\end{equation}
Here, 
\begin{equation}
	\langle j_\lambda(q) \rangle_{ELSj} = \int u_{LS}(r) j_\lambda(qr) R_{lj}(r) r^2 \dd r
\end{equation}
is the weighted radial wave function overlap \cite{N_v441_i_p486,M_v43_i9_p971} with the radial wave function $R_j(r)$ of the initial state, the radial wave function $u_{LS}(r)$ of the final state\footnote{This is to be understood as the radial wave function of the projection of the (delocalized) final Bloch state wave onto an $LS$ state centered at the scattering center, e.g., a muffin-tin state.}, and the spherical Bessel function $j_\lambda$. The $\sum_{\vec{k}n} D_{LMS}^{\vec{k}n} \left( D_{L'M'S'}^{\vec{k}n} \right)^*$ (over a shell of constant energy) is the cross density of states (XDOS) and the $\scriptsize\wigner{\bullet}{\bullet}{\bullet}{\bullet}{\bullet}{\bullet}$ are Wigner 3j symbols.

While this choice of basis is very convenient as a starting point (as it is used, e.g., in WIEN2k \cite{wien2k}), it is by no means the only or the optimal choice. This can be seen by collecting terms depending on $\vec{q}$ and terms depending on $\vec{q}'$. With the abbreviations
\begin{equation}
\begin{aligned}
	\alpha ={}& (\lambda, \mu, L, S) \\
	\alpha' ={}& (\lambda', \mu', L', S') \\
	g_\alpha(\vec{q}) ={}& \frac{1}{q^2} Y_\lambda^\mu(\vec{q}) \langle j_{\lambda}(q) \rangle_{ELSj} \\
	\Xi_{\alpha\alpha'} ={}& -\frac{\pi e^4}{\epsilon_0^2} (2l+1) (2j+1) \sum_{mm'}\sum_{MM'} \\
	& \ii^{\lambda-\lambda'} \sqrt{(2\lambda+1)(2\lambda'+1)(2L+1)(2L'+1)} \\
	& \wigner{l}{0}{\lambda}{0}{L}{0} \wigner{l}{0}{\lambda'}{0}{L'}{0} \\
	& \wigner{l}{-m}{\lambda}{\mu}{L}{M} \wigner{l}{-m'}{\lambda'}{\mu'}{L'}{M'} \\
	& \sum_{j_z} (-1)^{m+m'} \wigner{l}{m}{\frac{1}{2}}{S}{j}{-j_z} \wigner{l}{m'}{\frac{1}{2}}{S'}{j}{-j_z} \\
	& \sum_{\vec{k}n} D_{LMS}^{\vec{k}n} \left( D_{L'M'S'}^{\vec{k}n} \right)^* \delta(E + E_{nl\kappa} - E_{\vec{k}n}),
\end{aligned}
\end{equation}
the MDFF can be rewritten as
\begin{equation}
\label{eq:quadratic_form_1}
\begin{aligned}
	S(\vec{q}, \vec{q}') &= \sum_{\alpha\alpha'} g_{\alpha}(\vec{q})^\dag \Xi_{\alpha\alpha'} g_{\alpha'}(\vec{q}') \\
	&= \vec{g}(\vec{q})^\dag \cdot \mat{\Xi} \cdot \vec{g}(\vec{q}'),
\end{aligned}
\end{equation}
where the matrix $\mat{\Xi}$ collects all $\vec{q}, \vec{q}'$ independent terms and can be computed in a straight-forward way once the XDOS is known (e.g., from DFT calculations; note that this is a property of the target alone, and therefore has to be calculated only once). The $\vec{g}$, in turn, can be interpreted as a vector of functions. Eq.~\ref{eq:quadratic_form_1} is a well-known quadratic form and an extension of the often-used simple dipole approximation $S(\vec{q}, \vec{q}') = \vec{q} \cdot \mat{A} \cdot \vec{q}'$ \cite{PhysRevB.54.3861,U_v108_i5_p433,U_v111_i_p1163,PRB_v84_i_p64444} to arbitrary momentum transfers and multipole orders. In particular, it is noteworthy that $\mat{\Xi}$ is hermitian (as is shown in \ref{sec:hermiticity_xi}).

With the default settings, WIEN2k produces data with $0 \le L, L' \le 3$. When including transitions up to (and including) quadrupole order ($\lambda = 2$), $\mat{\Xi}$ is a $72 \times 72$ matrix, resulting in up to 5184 terms in the MDFF that in principle would all have to be handled separately. In practice, some of the entries vanish due to selection rules, while for some others the hermiticity of $\mat{\Xi}$ can be exploited. Still, many off-diagonal elements generally remain. These off-diagonal elements imply correlations between the basis vectors \cite{M_v31_i4_p333} and hence represent additional information (e.g., symmetries) about the underlying system that can be used to simplify the problem.

To exploit this additional information, one can insert a unitary matrix $\mat{U}$ in the following way:
\begin{equation}
	S(\vec{q}, \vec{q}') = \vec{g}(\vec{q})^\dag \cdot \mat{U}^\dag \mat{U} \cdot \mat{\Xi} \cdot \mat{U}^\dag \mat{U} \cdot \vec{g}(\vec{q}').
\end{equation}
Since for any hermitian matrix, a unitary matrix exists such that $\mat{U}\mat{\Xi}\mat{U}^\dag$ is a diagonal matrix $\mat{D}$, one only has to find such a $\mat{U}$. This is straight forward using, e.g., eigenvalue solvers, a singular value decomposition, or a Schur decomposition. With the abbreviation $\tilde{\vec{g}}(\vec{q}) = \mat{U}\cdot\vec{g}(\vec{q})$, the MDFF becomes
\begin{equation}
	S(\vec{q}, \vec{q}') = \tilde{\vec{g}}(\vec{q})^\dag \cdot \mat{D} \cdot \tilde{\vec{g}}(\vec{q}').
\end{equation}

In terms of quadratic forms, the transformation $\mat{U}$ is a principal axis transformation. In quantum mechanical terms, it is a basis transformation into the eigensystem of the MDFF. In essence, it recovers the ``physical'' basis of independent --- i.e., uncorrelated because of vanishing off-diagonal terms --- transitions.\footnote{Note that, depending on the EELS-edge and multipole orders considered, it may or may not be possible to determine the ``physical'' basis $\ket{F}$ of the \emph{target} from these transitions. Considering, e.g., a dipole-allowed transition from an initial p state to a final d state, one has only three transition elements ($\mu \in \{-1, 0, 1\}$), but 5 final states. Under these circumstances, not all information about the final states can be probed, unless one takes into account other multipole orders or final states.} With the default settings of WIEN2k, this means that the problem was reduced from at most 5184 terms to at most 72 terms. This is related directly to the internal crystal symmetries that are taken into account in WIEN2k. The information theoretical aspect of the effect (general) symmetry groups have on inelastic electron scattering has recently been studied by \citet{Dwyer2012}.

In practical applications, knowing the ``physical'' basis is important for understanding how the excitation process works. A priori, it is not clear how the independent transitions look like. One could have, e.g., a $p_x$-like transition as one would expect from $s \to p$ transitions, or a coherent superposition like a vortex-like transition similar to $(p_x - \ii p_y)/\sqrt{2}$, or something like an $s$-$p$ hybrid transition, or something even more complicated. Of course, all this information is present in the $\mat{\Xi}$ matrix, but it is not readily accessible in general. However, knowing it is very important when planning experiments (e.g., for knowing the diffraction angles at which to place an objective aperture, or to interpret recorded images in terms of these transitions and possibly different involved final states of the target). For this, the diagonalization can help as it produces exactly these uncorrelated transitions.

In numerical simulations, it is usually beneficial to work with the rMDFF as it can be multiplied directly onto the incident density matrix. Since the rMDFF is related to the MDFF by a Fourier transformation and the $\vec{q}$ and $\vec{q}'$ dependencies have been decoupled, the rMDFF simply reads
\begin{equation}
	S(\vec{r}, \vec{r}') = \tilde{\vec{g}}(\vec{r})^\dag \cdot \mat{D} \cdot \tilde{\vec{g}}(\vec{r}')
\end{equation}
with the same matrix $\mat{D}$ as for the MDFF and $\tilde{\vec{g}}(\vec{r}) = FT_{\vec{q}}[ \tilde{\vec{g}}(\vec{q}) ]$. By renormalizing the $\tilde{\vec{g}}$ such that $\bar{g}_\alpha(\vec{r}) := \sqrt{D_{\alpha\alpha}} \tilde{g}_\alpha(\vec{r})$, the MDFF can be further simplified to
\begin{equation}
	S(\vec{r}, \vec{r}') = \bar{\vec{g}}(\vec{r})^\dag \cdot \bar{\vec{g}}(\vec{r}') = \sum_\alpha \bar{g}_\alpha(\vec{r})^* \bar{g}_\alpha(\vec{r}')
\end{equation}

Hence, the reduced density matrix of the probe electron after the inelastic interaction in configuration space can be written as
\begin{equation}
	\mat{\rho}(\vec{r}, \vec{r}') = \sum_\alpha (\bar{g}_\alpha(\vec{r}) \phi(\vec{r}))^* \bar{g}_\alpha(\vec{r'}) \phi(\vec{r}'),
	\label{eq:rho_mat_r}
\end{equation}
where we wrote $\phi(\vec{r}) = \braket{\vec{r} | i}$ for the incident probe electron wave function. It is quite obvious that the diagonalization of the rMDFF has resulted in a pure state decomposition of the density operator $\matop{\rho} = \sum_\alpha \ketbra{\alpha}{\alpha}$. This is formally equivalent to the spectral decomposition of the cross-spectral density of quasi-monochromatic wave fields in optics \cite{JoMO_v57_i13_p1109}. However, contrary to the case in optics, we are dealing with the effects in electronic transitions. In particular, this results in the entanglement of the probe electron and the target, and thus in the necessity to construct the reduced density matrix of the probe beam.

Finally --- when measuring a real space image ---, the measurable intensity $I$ is given by\footnote{Here, an ideal lens system is assumed. Real lenses will reduce the level of detail transfered to the image, but do not change the coherence properties of the partial waves.}
\begin{equation}
	I(\vec{r}) = \mat{\rho}(\vec{r}, \vec{r}) = \sum_\alpha | \bar{g}_\alpha(\vec{r}) \phi(\vec{r}))|^2
	\label{eq:image_intensity}
\end{equation}

In the description above, elastic scattering of the probe beam after the inelastic scattering event has not been included for the sake of simplicity. Since each $\ket{\alpha}$ in itself is a pure state, it can be propagated elastically through the rest of the crystal with existing methods (e.g., the multislice approach \cite{Kirkland1998}) in a straight-forward way.

For the simulations in section \ref{sec:app_crystals}, the multislice approach was used first in order to propagate the incident beam through the crystal. At each atomic position, an inelastic interaction was simulated by calculating the (diagonal) reduced density matrix in eq.~\ref{eq:rho_mat_r}. The resulting independent pure states (or rather their corresponding wave functions) were then propagated through the rest of the crystal using the multislice approach again. Here, the diagonalization is of utmost importance as the number of multislice steps to perform is $\mathcal{O}(N L^2)$ where $N$ is the number of non-negligible terms in the MDFF (which can be reduced significantly by the diagonalization procedure) and $L$ is the number of layers (which is fixed by the geometry).

Obviously, the final image must be independent of the basis in which the inelastic scattering is described. However, using the diagonalization method presented here, the number of terms to calculate and hence the numerical complexity can be reduced considerably. In particular, it must be emphasized that this diagonalization has to be done only once, as a preprocessing step, but reduces the number of scattered wave function in each slice of the multislice calculation. Hence, it has a huge impact on the overall computation time.


\section{Applications}

In this section, we will apply the method outlined above to some model systems. The first two are included for didactic reasons as they demonstrate that the new method is consistent with previous findings. The third one is a more complicated real system.

\subsection{Single atom}

For single, individual atoms, all final states are independent of one another and hence uncorrelated. In addition, in the absence of a (strong) external magnetic field, states with the same $L$, but different $M$ or $S$ can be considered degenerate. For the XDOS, this means
\begin{equation}
	\sum_{\vec{k}n} D_{LMS}^{\vec{k}n} \left( D_{L'M'S'}^{\vec{k}n} \right)^* = D_{L} \delta_{LL'} \delta_{MM'} \delta_{SS'}.
\end{equation}
In the case of no spin-polarization, the $\langle j_\lambda \rangle$ also do not depend on $S$, the sum over $S,S'$ can be carried out, and the spin-dependence of $\alpha$ can be dropped.
Using the orthogonality relations for the Wigner 3j symbols,
a short calculation then yields
\begin{equation}
	\Xi_{\alpha\alpha'} = 4\pi (2L+1)(2j+1) \wigner{l}{0}{\lambda}{0}{L}{0}^2 D_L \delta_{\lambda\lambda'} \delta_{\mu\mu'} \delta_{LL'},
\end{equation}
Hence, for single, isolated atoms, $\mat{\Xi}$ is diagonal in the $(\lambda, \mu, L, S)$ basis, i.e., no correlations occur, even in the untransformed basis. More precisely, for the case discussed later, it is $2(2\lambda+1)$ fold degenerate in $\mu$ and $S$ (but not in $L$, in general). This is intimately connected to the fact that $\mat{\Xi}$ commutes with the rotation group of the single atom, and the $\lambda$\textsuperscript{th} irreducible representation of the full infinitesimal rotation group is $2(2 \lambda+1)$ dimensional. 


The angular dependence in the image is influenced only by the (azimuthal part of the) spherical harmonics\footnote{Assuming elastic scattering effects are negligible.}. Since all off-diagonal terms ($\mu \ne \mu'$) vanish and all $\mu$ have the same weighting, the incoherent summation over different $\ket{\alpha}$ in eq.~\ref{eq:image_intensity} gives terms of the form $\sum_\mu Y_\lambda^\mu (\vec{q}/q)^*Y_\lambda^\mu (\vec{q}/q)$. Since $q_z$ is given by the energy-loss and is constant \cite{U_v111_i_p1163}, this can be rewritten as $\sum_\mu Y_\lambda^\mu (\theta, \phi)^*Y_\lambda^\mu (\theta, \phi')$ with constant $\theta$. Fourier transformation into real space transforms $\exp(\ii\mu\phi)$ into $\exp(\ii\mu\phi)$ up to a phase factor \cite{IToPAaMI_v31_i_p1715}. So, measurements in real space (which imply $\phi=\phi'$) correspond to $\sum_\mu |Y_\lambda^\mu (\theta, \phi)|^2 = (2\lambda+1)/(4\pi)$ which is constant. This gives rise to circular intensity profiles regardless of the symmetries of the target's initial or final states.


\subsection{Energy-loss magnetic chiral dichroism}

Since its discovery in 2006 \cite{N_v441_i_p486}, interest in the energy loss magnetic chiral dichroism (EMCD) technique has been growing steadily. Using EMCD, one can determine the magnetic properties of the sample \cite{PRB_v76_i6_p60408}, similar to the X-ray magnetic circular dichroism which is a standard method in the synchrotron. The pure state decomposition approach outlined here can also be applied to EMCD.

For the sake of simplicity, we assume here a fully spin-polarized ($\delta_{S\frac{1}{2}}\delta_{S'\frac{1}{2}}$) dipole-allowed ($\lambda=\lambda'=1$) transition from an initial p ($l=1$) to a final d ($L=L'=2$) state, as is the dominant contribution to the L-edge in most common magnetic materials. In addition, we assume that states with same $L$, but different $M$ are (mostly) degenerate, as in the isolated atom case. Hence, the XDOS reads
\begin{equation}
	\sum_{\vec{k}n} D_{LMS}^{\vec{k}n} \left( D_{L'M'S'}^{\vec{k}n} \right)^* = D_2 \delta_{L2}\delta_{L'2} \delta_{MM'} \delta_{S\frac{1}{2}}\delta_{S'\frac{1}{2}}.
\end{equation}
Under these assumptions, $\mat{\Xi}$ becomes a $3 \times 3$ matrix for both $j=1/2$ (corresponding to the L$_2$ edge) and $j = 3/2$ (corresponding to the L$_3$ edge):
\begin{equation}
\begin{aligned}
	\mat{\Xi}_{j=1/2} &=
	\frac{4\pi D_2}{3}
	\begin{pmatrix}
		1 & 0 & 0 \\
		0 & 2 & 0 \\
		0 & 0 & 3
	\end{pmatrix} \\
	\mat{\Xi}_{j=3/2} &=
	\frac{4\pi D_2}{3}
	\begin{pmatrix}
		5 & 0 & 0 \\
		0 & 4 & 0 \\
		0 & 0 & 3
	\end{pmatrix}
\end{aligned}
\end{equation}
As in the single-atom case, $\mat{\Xi}$ is already diagonal in the spherical harmonics basis. Here, however, different $\mu$ have different weights. This means that transforming to any other basis will introduce off-diagonal elements (only the identity matrix is invariant under rotations). Hence, the spherical harmonics basis is the only ``physical'' basis for EMCD.

Moreover, the $\mat{\Xi}$ matrices given above can be interpreted as a homogeneous average signal on which the $\mu$-de\-pen\-dent EMCD signal is superimposed:
\begin{equation}
\begin{aligned}
	\mat{\Xi}_{j=1/2} &=
	\frac{4\pi D_2}{3} \left[
	\begin{pmatrix}
		2 & 0 & 0 \\
		0 & 2 & 0 \\
		0 & 0 & 2
	\end{pmatrix} +
	\begin{pmatrix}
		-1 & 0 & 0 \\
		0 & 0 & 0 \\
		0 & 0 & 1
	\end{pmatrix}
	\right] \\
	\mat{\Xi}_{j=3/2} &=
	\frac{4\pi D_2}{3}
	\left[
	\begin{pmatrix}
		4 & 0 & 0 \\
		0 & 4 & 0 \\
		0 & 0 & 4
	\end{pmatrix} +
	\begin{pmatrix}
		1 & 0 & 0 \\
		0 & 0 & 0 \\
		0 & 0 & -1
	\end{pmatrix}
	\right]
\end{aligned}
\end{equation}
This immediately shows two features common to EELS and EMCD. On the one hand, the homogeneous average signal exhibits the typical, statistical 1:2 intensity ratio of the L$_2$:L$_3$ edges. On the other hand, the EMCD signal (the absolute magnitude of which is independent of $j$ in this simple case) shows the typical sign reversal between L$_2$ and L$_3$ edges.

\subsection{Crystals}
\label{sec:app_crystals}

In crystals, the situation is more complicated and simple toy-models are insufficient to grasp them completely. Hence, one needs sophisticated calculations to derive the XDOS that take into account the full crystal structure. 	 \cite{Nelhiebel1999,Jorissen2007}.\footnote{The complete investigation of the effects of crystal symmetries on the XDOS and $\mat{\Xi}$ is beyond the scope of this work.}

\begin{figure}
	\includegraphics{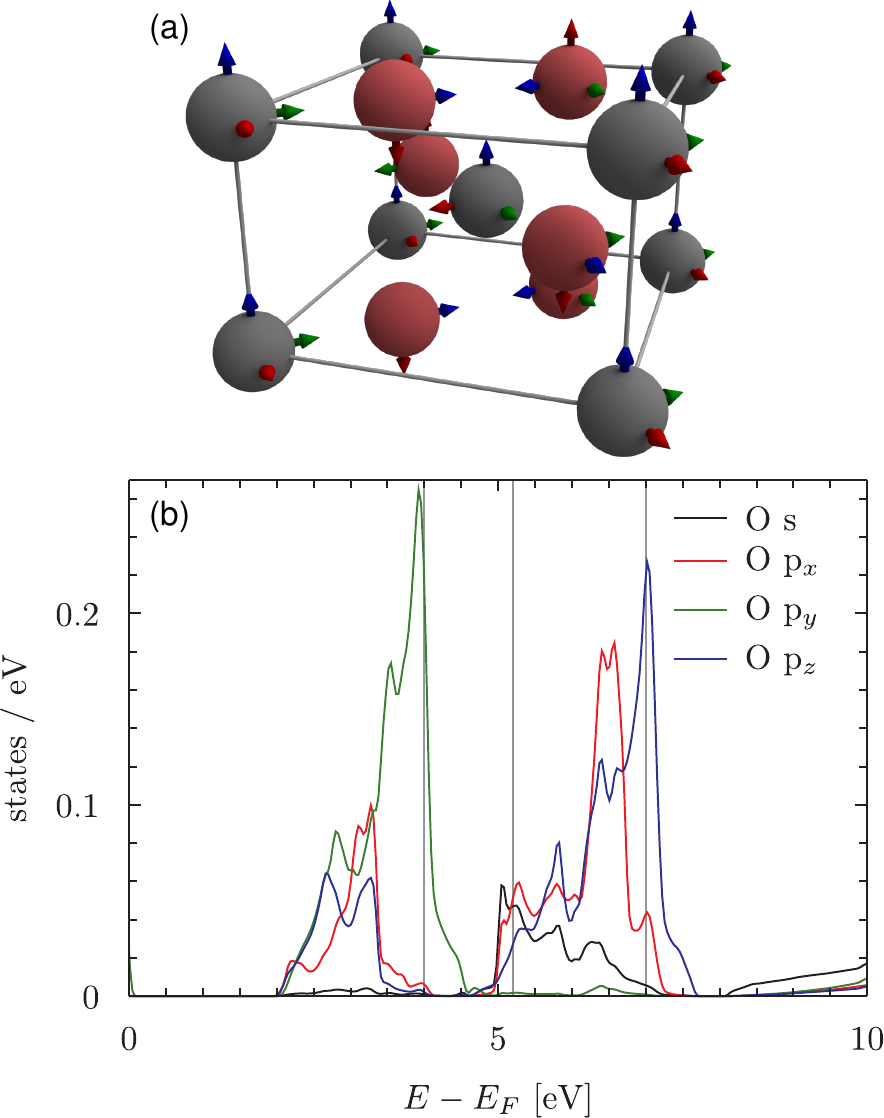}
	\caption{(a) schematic of the unit cell for Rutile (gray: Ti, red: O). The lattice constants are $a=\unit{4.594}{\angstrom}$ and $c=\unit{2.958}{\angstrom}$. The arrows show the symmetry-adapted local coordinate systems (red:~x, green:~y, blue:~z). (b) pDOS of the oxygen states as calculated by WIEN2k. The gray bars show energies used for simulations in this work.}
	\label{fig:rutile_wien2k}
\end{figure}

Hence, we will use the oxygen K-edge of Rutile (TiO$_2$), a tetragonal system, as test case in this work. Fig.~\ref{fig:rutile_wien2k}a shows a schematic of the unit cell, while fig.~\ref{fig:rutile_wien2k}b shows the partial density of states (pDOS) for oxygen as calculated by WIEN2k. From it, the lifting of the degeneracy of the different p~orbitals is already evident.

For this system, WIEN2k produces 89 non-negligible\footnote{Here, elements are considered non-negligible if they are larger than {1\textperthousand} of the largest element.} XDOS components at $E_F + \unit{4}{\electronvolt}$ in the spin-unpolarized case, whereas at $E_F + \unit{7}{\electronvolt}$, it produces 100 non-negligible elements. In the simplest case of taking into account only dipole-allowed transitions ($\lambda=\lambda'=1$), the $3 \times 3$ matrix $\mat{\Xi}$ has 5 non-vanishing entries, which are reduced to 3 after diagonalization.

\begin{figure}
	\includegraphics{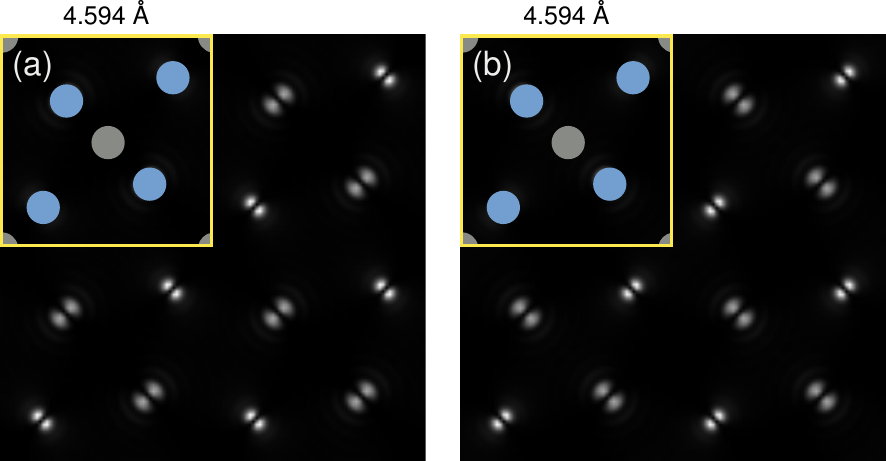}
	\caption{Real-space intensity of the exit wave after propagation of an incident plane wave through a one unit-cell thick crystal oriented in \hkl[0 0 1] zone axis at \unit{200}{\kilo\volt} acceleration voltage. Only dipole-allowed transitions were taken into account. (a) shows the image at an energy loss of $E_F + \unit{4}{\electronvolt}$, whereas (b) shows the image at an energy loss of $E_F + \unit{7}{\electronvolt}$. The inset shows the projected unit cell with Ti atoms in gray and O atoms in blue.}
	\label{fig:rutile-ideal}
\end{figure}

Fig.~\ref{fig:rutile-ideal}a shows the simulated exit wave function intensities (corresponding to an ideal lens system) for a single unit cell after an energy loss of $E_F + \unit{4}{\electronvolt}$. The p type character of the transitions is clearly visible to be pointing in the directions of the green axes in fig.~\ref{fig:rutile_wien2k}a. Because of the simple, spherically symmetric 1s nature of the initial state of the oxygen, these p$_\text{y}$ type transitions correspond directly to p$_\text{y}$ type final states of the oxygen.

Likewise, fig.~\ref{fig:rutile-ideal}b shows the simulated exit wave function intensities at an energy loss of $E_F + \unit{7}{\electronvolt}$. There, p type transitions pointing along the blue axes in fig.~\ref{fig:rutile_wien2k}a corresponding to oxygen p$_\text{z}$ orbitals are clearly visible. Naturally, they are rotated by \unit{90}{\degree} with respect to the p$_\text{y}$ orbitals.

\begin{figure}
	\centering
	\includegraphics{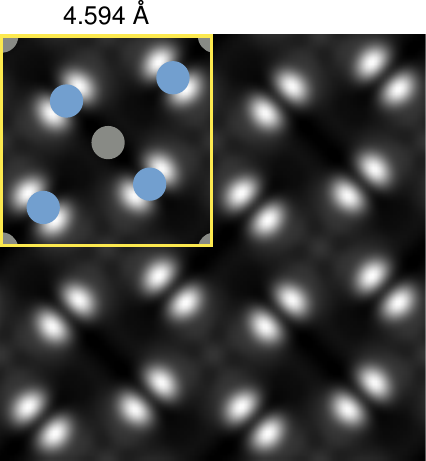}
	\caption{Real-space intensity of the exit wave after propagation of an incident plane wave through a \unit{10}{\nano\meter} thick crystal oriented in \hkl[0 0 1] zone axis at \unit{200}{\kilo\volt} acceleration voltage. A \unit{24}{\milli\radian} objective aperture was used. Only dipole-allowed transitions were taken into account. The image is taken at an energy loss of $E_F + \unit{4}{\electronvolt}$. The inset shows the projected unit cell with Ti atoms in gray and O atoms in blue.}
	\label{fig:rutile-real}
\end{figure}

Fig.~\ref{fig:rutile-real} shows the situation for a \unit{10}{\nano\meter} thick crystal and an objective aperture of \unit{24}{\milli\radian}. For these calculations, elastic scattering both before and after the inelastic interaction was taken into account using the multislice approach \cite{Kirkland1998}. This demonstrates that these results are not only of theoretical interest, but should be measurable in real instruments. The contrast can be estimated to be \unit{96}{\%} (after subtraction of the pre-edge background, e.g., using the three-window method).

\begin{figure}
	\centering
	\includegraphics{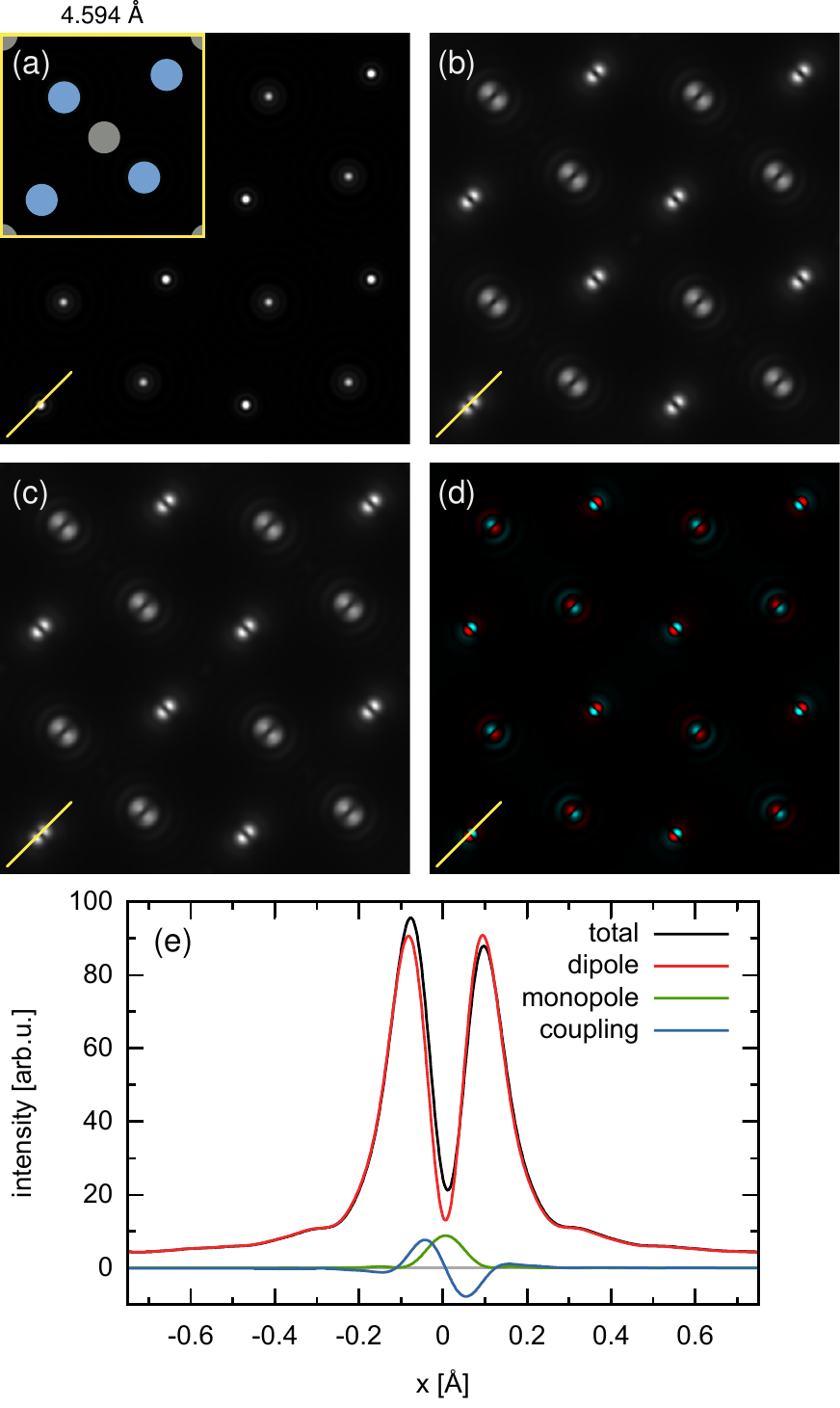}
	\caption{(a) - (d) Real-space intensities of the exit wave after propagation of an incident plane wave through a one unit cell thick crystal oriented in \hkl[0 0 1] zone axis at \unit{200}{\kilo\volt} acceleration voltage. (a) shows only monopole contributions (contrast-enhanced by a factor of 15), (b) shows only dipole contributions, (c) shows the total intensity, and (d) shows the coupling contribution (contrast enhanced by a factor of 15; red indicates positive values while cyan represents negative ones). The images are taken at an energy loss of $E_F + \unit{5.2}{\electronvolt}$. The inset shows the projected unit cell with Ti atoms in gray and O atoms in blue. (e) Traces of the different contributions. The places of the trace are marked by yellow lines}
	\label{fig:lambda-coupling}
\end{figure}

More importantly, non-dipole transitions can easily be taken into account as well. For the oxygen K-edge, the most relevant non-dipole transition is the monopole transition from the 1s state to final states with s symmetry\footnote{The pDOS for d states as produced by WIEN2k that would be accessible by quadrupole-allowed transitions is negligibly small.}. Here, the number of non-negligible terms was reduced from 10 to 4. Fig.~\ref{fig:lambda-coupling} compares the intensities and images of monopole-allowed transitions, dipole-allowed transitions, and the coupling term between the two. A similar effect was predicted recently for X-ray absorption spectrometry~\cite{PRB_v85_i_p115136}.

Interestingly, the coupling term gives intensity variations of about $\unit{\pm 10}{\%}$ of the dipole-allowed transitions, which is comparable to the monopole transitions. Because the coupling term has a sign, this means that at some positions it roughly cancels the monopole contributions, whereas at other positions it can even double it. This acts in a way very similar to s-p hybridization, yielding an asymmetric image. 

\section{Conclusion and Outlook}

In this work, we demonstrated a method to diagonalize both the MDFF and the rMDFF leading to a pure state decomposition of the density operator. This was shown to yield obvious numerical advantages by reducing the number of terms to include in image calculations. Moreover, the diagonalization leads to a new set of basis vectors that are helpful to elucidate the physics underlying the scattering process.

The new pure state decomposition method was applied to the isolated atom case, EMCD, and a Rutile crystal to show its versatility. In particular, the isolated atom and the EMCD cases could be treated analytically, giving results from which important properties such as the L$_2$:L$_3$ ratio or the sign reversal of the EMCD effect could be seen immediately.

For the Rutile crystal, it was shown that with latest-generation TEMs, it should be possible to directly map atomic orbitals, e.g., using energy filtered TEM (EFTEM) with high spatial resolution. However, contrary to the common assumption that non-dipole transitions are unimportant, it was shown that the monopole as well as the monopole-dipole coupling terms can change the signal measurably. The necessary condition to achieve this is to have a system with sufficiently low symmetry (otherwise off-diagonal terms vanish due to symmetry considerations~\cite{Nelhiebel1999,Jorissen2007}).

Based on this, the situation for other low-symmetry situation should be the same. Hence, this new technique gives rise to exciting new possibilities like directly studying the electronic structure of defects (see, e.g., \cite{NM_v10_i3_p209}), interfaces, or other low-symmetry objects.

\section*{Acknowledgements}

The authors acknowledge financial support by the Austrian Science Fund (FWF) under grant number I543-N20. They also want to express their gratitude to Walid Hetaba for fruitful discussions about and help using WIEN2k.

\appendix

\section{Hermiticity of \protect{$\Xi$}}
\label{sec:hermiticity_xi}

For $\mat{\Xi}$ to be hermitian, the equation
\begin{equation}
	\Xi_{\alpha'\alpha} = \Xi_{\alpha\alpha'}^*
\end{equation}
must hold.
\begin{equation}
\begin{aligned}
	\Xi_{\alpha'\alpha} =& 4\pi (2l+1) (2j+1) \sum_{m'm}\sum_{M'M} \delta(E + E_{nl\kappa} - E_{\vec{k}n})\\
	& \sqrt{(2\lambda'+1)(2\lambda+1)(2L'+1)(2L+1)} \\
	& \wigner{l}{0}{\lambda'}{0}{L'}{0} \wigner{l}{0}{\lambda}{0}{L}{0} \\
	& \wigner{l}{-m'}{\lambda'}{\mu'}{L'}{M'} \wigner{l}{-m}{\lambda}{\mu}{L}{M} \\
	& \sum_{j_z} (-1)^{m'+m} \wigner{l}{m'}{\frac{1}{2}}{S'}{j}{-j_z} \wigner{l}{m}{\frac{1}{2}}{S}{j}{-j_z} \\
	& \ii^{\lambda'-\lambda}\sum_{\vec{k}n} D_{L'M'S'}^{\vec{k}n} \left( D_{LMS}^{\vec{k}n} \right)^*  \\
	=& 4\pi (2l+1) (2j+1) \sum_{mm'}\sum_{MM'} \delta(E + E_{nl\kappa} - E_{\vec{k}n}) \\
	& \sqrt{(2\lambda+1)(2\lambda'+1)(2L+1)(2L'+1)} \\
	& \wigner{l}{0}{\lambda}{0}{L}{0} \wigner{l}{0}{\lambda'}{0}{L'}{0} \\
	& \wigner{l}{-m}{\lambda}{\mu}{L}{M} \wigner{l}{-m'}{\lambda'}{\mu'}{L'}{M'} \\
	& \sum_{j_z} (-1)^{m+m'} \wigner{l}{m}{\frac{1}{2}}{S}{j}{-j_z} \wigner{l}{m'}{\frac{1}{2}}{S'}{j}{-j_z} \\
	& \left[ \ii^{\lambda-\lambda'} \sum_{\vec{k}n} D_{LMS}^{\vec{k}n} \left( D_{L'M'S'}^{\vec{k}n} \right)^* \right]^* \\
	=& \Xi_{\alpha\alpha'}^*
\end{aligned}
\end{equation}

\section*{Bibliography}

\bibliography{mdff-factorization}

\end{document}